\documentclass[twocolumn,8pt]{aastex631}

\usepackage{graphicx,amsmath}
\usepackage{color}
\usepackage{booktabs}
\usepackage{dutchcal}
\usepackage{multirow}
\usepackage{amsmath}
\usepackage{cuted}

\newcommand{\adsurl}[1]{\href{#1}{ADS}}
\providecommand{\url}[1]{\href{#1}{#1}}

\newcommand{\be}{\begin{equation}}
\newcommand{\ee}{\end{equation}}
\newcommand{\bea}{\begin{eqnarray}}
\newcommand{\eea}{\end{eqnarray}}

\usepackage{color}
\usepackage{ifthen}
\newboolean{editorial}
\setboolean{editorial}{true}
\newcommand{\editorial}[2]{\ifthenelse{\boolean{editorial}}{\textcolor{red}{[\textsf{\textbf{{#1}}}: }\textcolor{blue}{\textsf{{#2}}}\textcolor{red}{]}}{}}

\shorttitle{Cosmic Variance on BNS merger rate density}
\shortauthors{Chen et al.}

\begin{document}

\title{On the Cosmic Variance of the Merger Rate Density of Binary Neutron Stars }

\correspondingauthor{Zhiwei Chen}
\email{chenzhiwei171@mails.ucas.ac.cn}

\author[0000-0001-7952-7945]{Zhiwei Chen}
\affiliation{National Astronomical Observatories, Chinese Academy of Sciences, 20A Datun Road, Beijing 100101, China}
\affiliation{School of Astronomy and Space Sciences, University of Chinese Academy of Sciences, 19A Yuquan Road, Beijing 100049, China}
\author[0000-0002-1310-4664]{Youjun Lu}
\affiliation{National Astronomical Observatories, Chinese Academy of Sciences, 20A Datun Road, Beijing 100101, China}
\affiliation{School of Astronomy and Space Sciences, University of Chinese Academy of Sciences, 19A Yuquan Road, Beijing 100049, China}
\author{Jie Wang}
\affiliation{National Astronomical Observatories, Chinese Academy of Sciences, 20A Datun Road, Beijing 100101, China}
\affiliation{School of Astronomy and Space Sciences, University of Chinese Academy of Sciences, 19A Yuquan Road, Beijing 100049, China}

\author{Zhen Jiang}
\affiliation{ Department of Astronomy, Tsinghua University, Beijing 100084, China }
\author{Qingbo Chu}
\affiliation{National Astronomical Observatories, Chinese Academy of Sciences, 20A Datun Road, Beijing 100101, China}
\affiliation{School of Astronomy and Space Sciences, University of Chinese Academy of Sciences, 19A Yuquan Road, Beijing 100049, China}
\author{Xianghao Ma}
\affiliation{National Astronomical Observatories, Chinese Academy of Sciences, 20A Datun Road, Beijing 100101, China}
\affiliation{School of Astronomy and Space Sciences, University of Chinese Academy of Sciences, 19A Yuquan Road, Beijing 100049, China}

\begin{abstract}
The cosmic variance on the star formation history may lead to bias to the merger rate density estimation of binary neutron star (BNS) mergers by the compact binary population synthesis. In this paper, we take the advantage of the large boxsize of the Millennium Simulation combined with the semi-analytic galaxy formation model GABE, and the  parameterized population binary star evolution (BSE) model to examine how much effect will the cosmic variance introduce on the estimation of merger rate density of BNS mergers. 
We find that for sub-box size of $100\rm Mpc$ and $200\rm Mpc$, the variance of merger rate density $\sigma_{\rm R}/\rm R$ at different redshift is about $23\%-35\%$ and $13\%-20\%$ respectively. On one hand, as for the variance of the detection rate on BNS mergers with current  LIGO-Virgo-KAGRA (LVK) detector network, this value is very small $\lesssim 10\%$, 
which indicates ignoring the cosmic variance is reasonable for estimating the merger rate density from current LVK observation. On the other hand, with next-generation gravitational wave detectors, it is possible to localize BNS mergers within sub-boxes possessing length of $\rm 40 Mpc$ for source redshift $z_{s}<0.2$. In such a small box, the cosmic variance of the merger rate density is significant, i.e., the value of $\sigma_{\rm R}/\rm R$ is about $\sim 55\%$.  This hints that estimating the merger rate density of BNS in
different sky areas may provide useful information
on the cosmic variance.
\end{abstract}


\section{Introduction}
\label{sec:intro}
The first detection of gravitational wave (GW) emitted by the binary neutron star (BNS) merger
GW170817 and its electromagnetic \textbf{(EM)} counterpart signals marks the beginning of a new era of 
multi-messenger astronomy \citep[e.g., ][]{2017ApJ...848L..12A, 2017ApJ...848L..13A, 2017Sci...358.1556C, 2017Natur.551...67P,2018ApJ...858L..15D, 2019MNRAS.489L..91C, 2019PhRvX...9a1001A}. The associated EM counterpart can provide significant information on the physical properties of BNS mergers, for example, the tidal deformability and the equation of state constrained from the GW, afterglow, and kilonova signals \citep[e.g.,][]{2018MNRAS.479..588G, 2018ApJ...863...58X}. Moreover, with the redshift measurement of the EM counterpart and luminosity distance measurement from the GW signal, BNS mergers can be viewed as excellent standard sirens to constrain the cosmological parameters. For example, the Hubble constant can be constrained to a precision of $\sim 10\%$ by the data of GW170817, GRB170817A, and AT 2017gfo \citep{2021arXiv211103634T}. With the development of ground-based GW detectors, it is anticipated to observe many more GW170817-like objects in the near future, which may constrain $\rm H_0$ to $\sim 1\%$ precision.

It is crucial to estimate the evolution of merger rate density of BNS mergers.   The merger rate density of the BNS mergers are dependent on the detailed physics embedded in the binary evolution procedure, including the common envelope ejection, Roche lobe mass transfer process, and the production of natal kick \citep[e.g., ][]{2022MNRAS.509.1557C}. For example, if the natal kick is large enough, the BNS may not form at the end of its progenitor life, which may decrease the merger rate density of BNS mergers significantly.  Another important recipe is the star-formation-rate (SFR) evolution of galaxies in our universe. The lower SFR will result in less compact binary formation and thus lower the merger rate density.  Therefore, it is possible 
to constrain such important physical recipes by comparing the measurement to the results given by compact binary population synthesis.

Cosmological numerical simulation has been widely used to estimate the merger rate density and predict the performance of ground-based GW detectors of double compact objects.
However, almost all the works done so far just simply take an average (global) of the resulted merger 
rate density in the full box of the hydrodynamical simulations  \citep[e.g.,][]{2012ApJ...759...52D, 2013ApJ...779...72D, 2015ApJ...806..263D, 2015ApJ...814...58D, 2018MNRAS.479.4391M,2022MNRAS.509.1557C}, such as Illustris-TNG \citep{2018MNRAS.475..648P} and EDGES \citep{2015MNRAS.446..521S},  limited by their small boxsize, normally $\sim300 \rm Mpc$, while ignoring the potential effect of cosmic variance in our local universe. This \textbf{choice} may lead to a bias to the merger rate density evolution estimation and therefore the detection rate of BNS mergers, especially in the O2 run of LIGO-Virgo-KAGRA (hereafter LVK) GW detector network \citep{2019PhRvX...9c1040A}, for its small angle-averaged BNS detection range ($\sim 96\rm Mpc$ for the LIGO-Livingston and $80\rm Mpc$ for the LIGO-Hanford). 

In this paper, we take the advantage of the large boxsize of the Millennium Simulation combined with the semi-analytic galaxy formation model GABE, and the parameterized population BSE model of our previous work to examine how much effect will the cosmic variance introduce on the estimation of merger rate density of BNS mergers. 
This paper is organized as follows. In Section~\ref{sec:simu}, we briefly introduce the semi-analytic galaxy formation model GABE and the key parameters of BSE models.  In Section~\ref{sec:results}, we present main results. Discussions and conclusions are given in Section~\ref{sec:con}. Throughout the paper, we adopt the cosmological parameters as $(h_0,\Omega_{\rm m},\Omega_\Lambda)=(0.68,0.31,0.69)$ \citep{Aghanim2020}.

\section{Simulation Recipes}
\label{sec:simu}

The semi-analytical galaxy formation model GABE and N-body numerical simulation Millennium \citep[e.g.,][]{2005Natur.435..629S, 2019RAA....19..151J} is adopted to estimate the cosmic merger rate density evolution of BNSs. The GABE model considers several important physical processes of galaxy formation, including the cosmic reionization, gas cooling with background radiation, star formation, supernova feedback, black hole growth, AGN feedback, and bar formation. As for the hierarchical growth of structure formation under the standard cold dark matter model, the tidal and ram-pressure stripping of hot gas, dynamical fraction, black hole growth and star bursts in mergers and bulge formation are taken into account. More detailed description about the galaxy formation and evolution in GABE can be seen in \citet{2019RAA....19..151J}.

The simulation box size is $685$\,Mpc and the mass resolution of dark matter is $1.2\times 10^9 M_{\odot}$, which allows us to generate a complete galaxy catalogue for galaxies more massive than $10^{8} M_{\odot}$. We discrete the full box into $216$ and $27$ sub-boxes with boxsize of $100$\,Mpc and $200$\,Mpc
respectively, which is about the twice the length of the comoving distance of the observed GW170817 ($\sim 50\rm Mpc$) and O2 BNS observation horizon ($\sim 90 \rm Mpc$). Note that we do not consider the length scale of O3-O4a run \citep{2021PhRvX..11b1053A,2023PhRvX..13d1039A}, for their large observation horizon, i.e., $\sim 150-200\rm Mpc$. 
In such a scale, the cosmic variance is small enough for one to neglect its impact.  

The merger rate density evolution of BNSs are mainly dependent on the metallicity and SFR density distribution across the universe, however, the dependency on metallicity is relatively weak.
Therefore, the diversity of detection rate of BNS among those sub-boxes is due to the difference of galaxy-formation history in different sub-boxes caused by cosmic variances. Then, for each sub-box, we calculate the evolution of metallicity and SFR for each galaxy through tracing back its formation history by finding out all the progenitors in the whole cosmic merger trees. 

\begin{figure*}
\centering
\includegraphics[width=1.0\columnwidth]{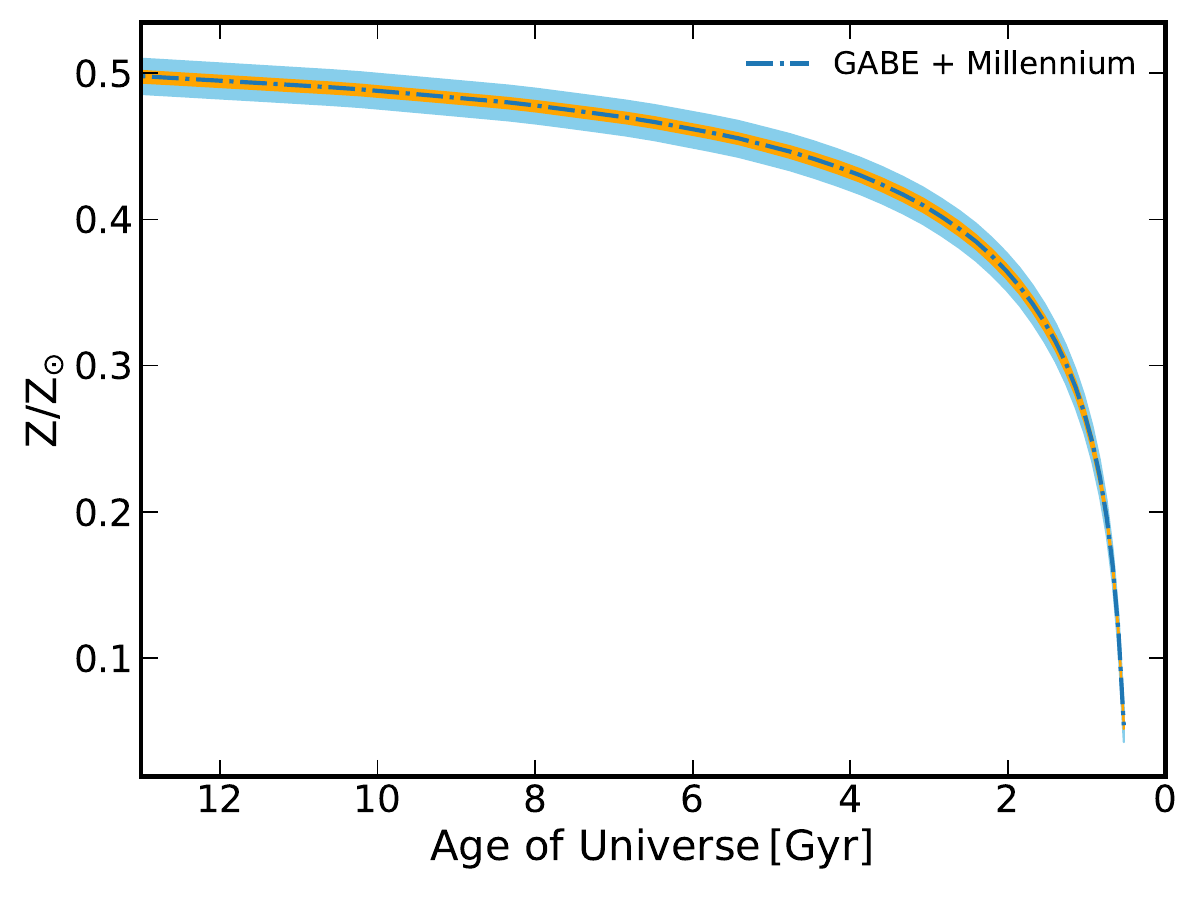}
\includegraphics[width=1.0\columnwidth]{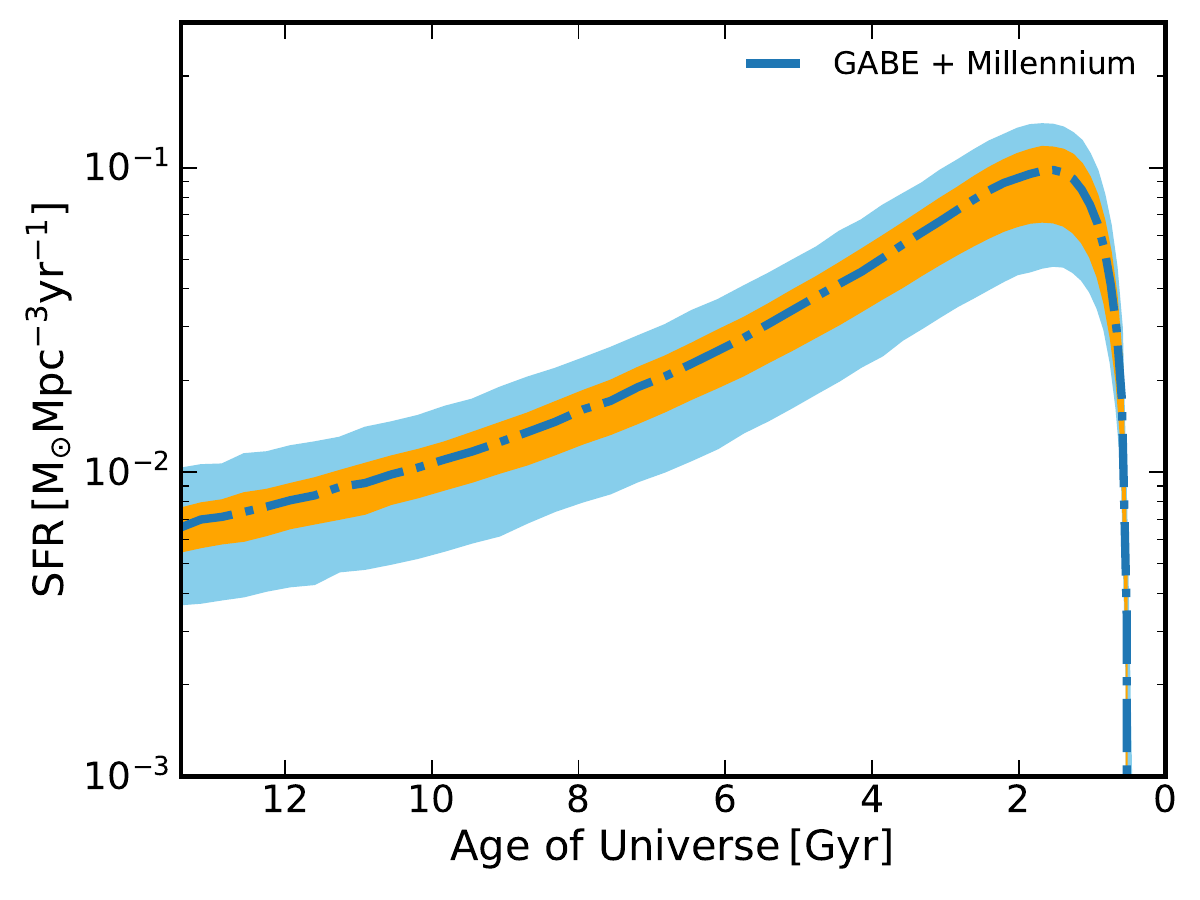}
\caption{The metallicity evolution (left panel) and star formation rate SFR density evolution (right panel) with redshift $z_{\rm s}$ from the GABE semi-analytical galaxy formation model together with Millenimum simulation. The blue dotted-dashed lines show the global results for both metallicity $Z$ and SFR. The \textbf{inner orange shaded} area represents for the scatter induced by cosmic variance when $N_{\rm subbox}=27$ and correspondingly $l_{\rm subbox}=200\rm Mpc$,  while for \textbf{outer blue shaded} area represents the scatter when $N_{\rm subbox}=216$ and correspondingly $l_{\rm subbox}=100\rm Mpc$. }
%
\label{fig:sfr}
\end{figure*}

Figure~\ref{fig:sfr} shows the metallicity $Z$ (left panel) and SFR evolution (right panel) with redshift $z_{\rm s}$ obtained from the semi-analytical galaxy formation model GABE and Millennium simulation. The blue dotted-dashed lines show the global (or the median) results for both metallicity $Z$ and SFR evolution, while the inner and outer shaded regions represent the scatter induced by the cosmic variance among $27$ sub-boxes with $l_{\rm subbox}=200\rm Mpc$ and $216$ sub-boxes with $l_{\rm subbox}=100\rm Mpc$, respectively. 
We note here that for metallicity, our result are slightly higher than those constrained by \citet{2016Natur.534..512B}, but almost the same as 
the results in \citet{2022MNRAS.509.1557C} obtained from EAGLE and Illustris-TNG. Moreover, one can see that the metallicity $Z$ and its evolution do not affect much by
the cosmic variance among different sub-boxes. However, for the SFR density, the cosmic variance can have a significant effect, for which the resulted SFR density may vary by a factor of $\sim 1.3-2$. 
The largest SFR produced in a subbox can be higher than that with the smallest SFR by a factor of more than $3$. Though it is currently not clear how much of an effect cosmic variance can introduce to the SFR density within the local volume \citep{2014ARA&A..52..415M},  observation of galaxy cluster survey shows that there  should be an underdensity in the matter distribution of about $\sim -30\pm 15\%$ in a region with a radius of about $100\rm Mpc$ \citep{2020A&A...633A..19B}. This may hint that the SFR density in local volume may be smaller than the average across the universe. Later, we shall see that the the cosmic variance affect the merger rate density of BNS mergers mainly via the scatter of SFR density.

\begin{table}
\caption{Model parameters of BSE models adopted, i.e., $\boldsymbol{\alpha10.\rm kb\beta0.9}$, $\boldsymbol{\alpha1.\rm kb\beta0.9}$ and $\boldsymbol{\gamma1.5\rm kb\beta0.9}$. Here the 1st column $\alpha/\gamma$ represent parameters of the $\alpha/\gamma$ formalism of the common envelope ejection process respectively. 
The 2nd column $\sigma_{\rm k}$ is the dispersion of the natal kick distribution, which is assumed to be Maxwellian and bimodal, with $\sigma_{\rm k}=190\rm  km\,s^{-1}$ and $30 \rm  km\,s^{-1}$, respectively. The third column $\beta$ represents the ratio of post-supernova to pre-supernova total system masses, which is set to be $0.9$ across these three models. More details about the BSE model can be seen in \citet{2022MNRAS.509.1557C}.}
\label{table:para} 
\centering
\begin{tabular}{lccc} 
\hline
\hline
Model  & $\alpha$ / $\gamma$ & $\sigma_{\rm k}(\rm km \,s^{-1})$ & $\beta$  \\
\hline
$\boldsymbol{\alpha10.\rm kb\beta0.9}$ & 10 & 190/30 & 0.9 \\
$\boldsymbol{\alpha1.\rm kb\beta0.9}$ & 1.0 & 190/30 & 0.9 \\
$\boldsymbol{\gamma1.5\rm kb\beta0.9}$ &
1.5 & 190/30 & 0.9 \\
\hline
\hline

\end{tabular}
\end{table}

We adopt the parameterized population models proposed in \citet{2022MNRAS.509.1557C} to mock the binary stellar evolution, including  $\boldsymbol{\alpha10.\rm kb\beta0.9}$, $\boldsymbol{\alpha1.\rm kb\beta0.9}$ and $\boldsymbol{\gamma1.5\rm kb\beta0.9}$. Among these models, the impacts of the common-envelope phase, natal kick, mass ejection during the secondary SN explosion, and metallicity are taken into account. 
The main differences between these three models are the choice of common-envelope description and the settings of model parameters. Here $\alpha$ and $\gamma$ denote for the conserved energy and conserved angular momentum description in the mass-transfer process of common-envelope respectively. 
In our previous work,  \citet{2022MNRAS.509.1557C} found that these three models rank the first three in the consistency check with the galactic BNS observations, by utilizing the Bayes factor methods. The key model parameters are listed in Table~\ref{table:para} and more details about  can be seen in \citet{2022MNRAS.509.1557C}.

\begin{figure}
\centering
\includegraphics[width=1.0\columnwidth]{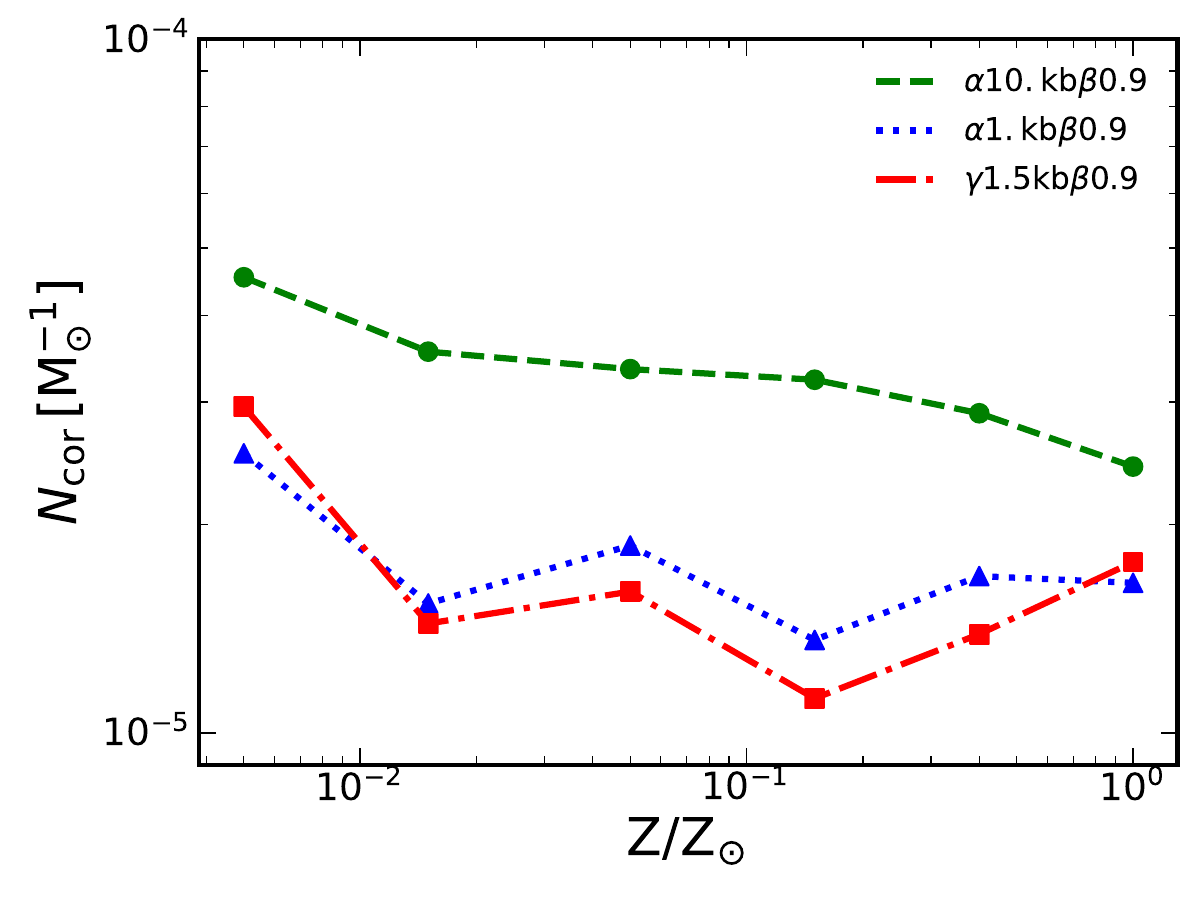}
\caption{
The relationship between number of mergering BNSs per unit mass ($N_{\rm cor}$) and the binary progenitor's metallicity (Z) simulated in \citet{2022MNRAS.509.1557C}. The green dashed, blue dotted, and red dotted-dashed lines show the results of different parameterized population models, i.e.,  $\boldsymbol{\alpha10.\rm kb\beta0.9}$, $\boldsymbol{\alpha1.\rm kb\beta0.9}$ and $\boldsymbol{\gamma1.5\rm kb\beta0.9}$, respectively.  
}
\label{fig:ncor}
\end{figure}

To reduce the cost of computing time, we do not implement the full BSE code into the GABE model together with Millenuim simulation for each sub-box, but rather apply the $N_{\rm cor}$ description as discussed in \citet{2018MNRAS.479.4391M} to estimate the merger rate density evolution of BNSs $R^{i}(z_s)$ in the $i$-th discrete sub-box as
\begin{equation}
R^{i}(z_s)=\int dt_{\rm d}P_t(t_{\rm d}) f_{\rm b}{\rm SFR^{i}}(z_{\rm b})\times N_{\rm cor}(Z^{i}\left(z_{\rm b})\right),
\label{eq:merger_rate}
\end{equation}
where $f_{\rm b}$ denotes the binary formation efficiency (normally $f_{\rm b}=0.5$), ${\rm SFR}^{i}(z_{\rm b})$ is the star-formation rate density at the binary formation redshift $z_{\rm b}$ (or time $t_{\rm b}$) in the $i$-th sub-box, and $N_{\rm cor}$ is the number of mergering BNSs per unit mass in the $i$-th sub-box, 
which is directly related to the star metallicity $Z^{i}(z_{\rm b})$ at $z_{\rm b}$. The term $t_{\rm d}(z_{\rm b})=\int_{z_{\rm s}}^{{z_{\rm b}}}\mid\frac{dt}{dz}\mid dz$ denotes the time delay of a BNS merger from its progenitor binary star formation time and $P_t(t_{\rm d})$ is its probability distribution, ranging from $10\rm Myr$ to a Hubble time $t_{\rm H}$. Here we argue that the above estimation based on this simplified description is solid for $N_{\rm cor}$ is only related to the metallicity $Z$ and different BSE models. 
Figure~\ref{fig:ncor} shows the dependence of $N_{\rm cor}$ as function of the binary progenitor's metallicity $Z$ simulated by \citet{2022MNRAS.509.1557C}. 
As seen from this figure that the value of $N_{\rm cor}$ can vary by a factor of $1-2$ considering of different metallicities in different BSE models. This is due to the small mass range of NSs, regardless of whether it is formed through iron core-collapse or electron-capture SNe.

\begin{figure}
\centering
\includegraphics[width=1.0\columnwidth]{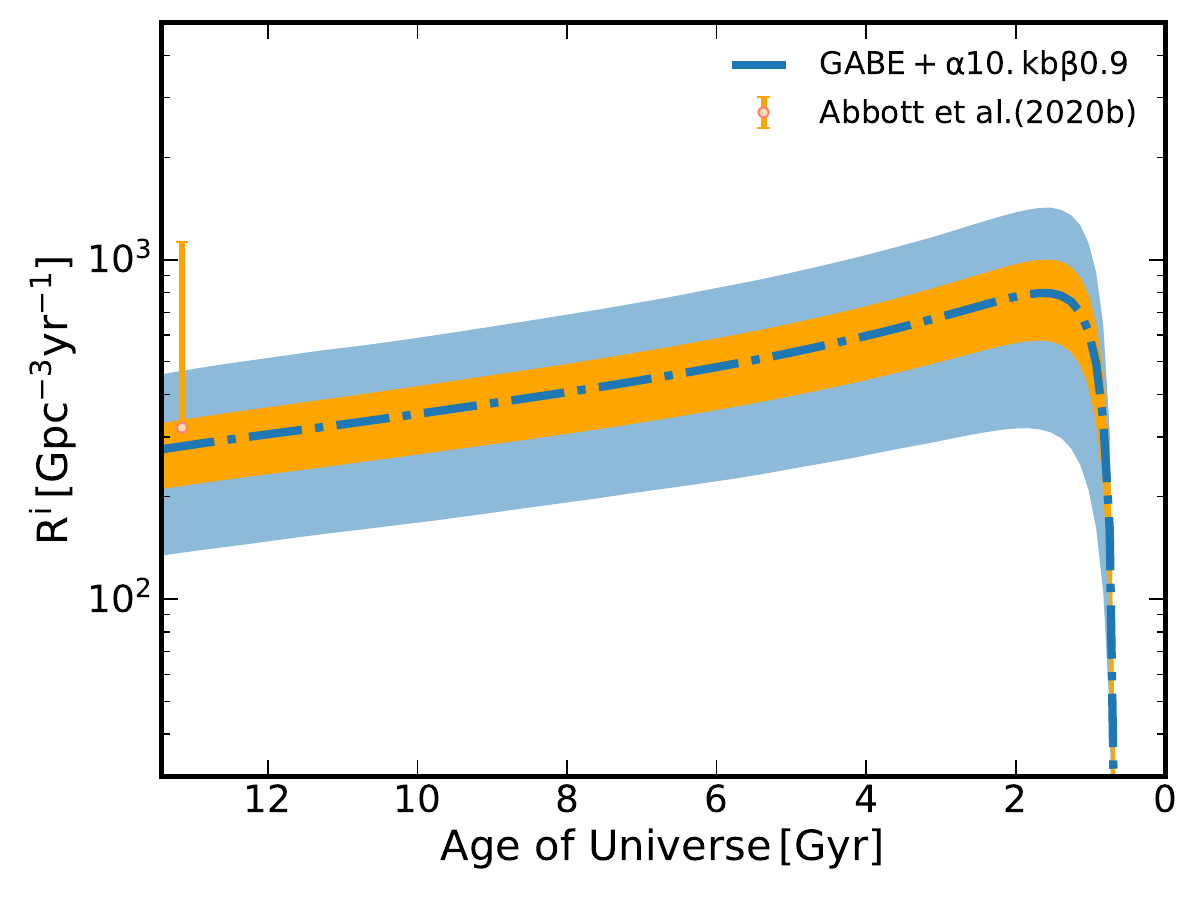}
\caption{
The comoving merger rate density evolution of BNS mergers $R^{i}$ as a function of the age of universe obtained by adopting the $\boldsymbol{\alpha10.\rm kb\beta0.9}$ BSE model. The blue dotted-dashed line shows the global results obtained by using the whole simulation box of \textbf{Millennium} simulation, and \textbf{the inner orange and outer  blue shaded regions} show the scatter of $R^{i}$ induced by the cosmic variance among the $27$ and $216$ discrete sub-boxes with size of $l_{\rm subbox}=100$\,Mpc and $200$\,Mpc, respectively. The orange dot with error bar marks the local merger rate density obtained from the GW detections of the two BNS mergers (GW170817 and GW190425) \citep{2020ApJ...896L..44A}, i.e., $R(z_{\rm s}=0)\sim 320_{-240}^{+490} \rm \,Gpc^{-3}\,yr^{-1}$. 
}
\label{fig:rm}
\end{figure}

\section{Results}
\label{sec:results}

We obtain the the merger rate density evolution $R^{i}$ of the BNS mergers in each sub-box as well as the global one from the whole box of Millnennium simulation by integrating Equation~\eqref{eq:merger_rate} over all the snapshots of our cosmological simulation, i.e., $z_{\rm s}\sim0-10$ for $64$ snapshots. Figure~\ref{fig:rm} shows the results of the merger rate density and its evolution by adopting the BSE model $\boldsymbol{\alpha10.\rm kb\beta0.9}$. 
 As seen from this figure, despite the slight bias of the median value ($R(z_{\rm s}=0)\sim 279 \rm Gpc^{-3}yr^{-1}$in our simulation), there is variation on $R(z_{\rm s})$ across the sub-boxes due to the cosmic variance. For example at the same redshift, the largest $R$ is almost twice the value of the smallest $R$ assuming $l_{\rm subbox}=100\rm Mpc$, for the SFR density in this sub-box is much higher.

\begin{figure}
\centering
\includegraphics[width=1.0\columnwidth]{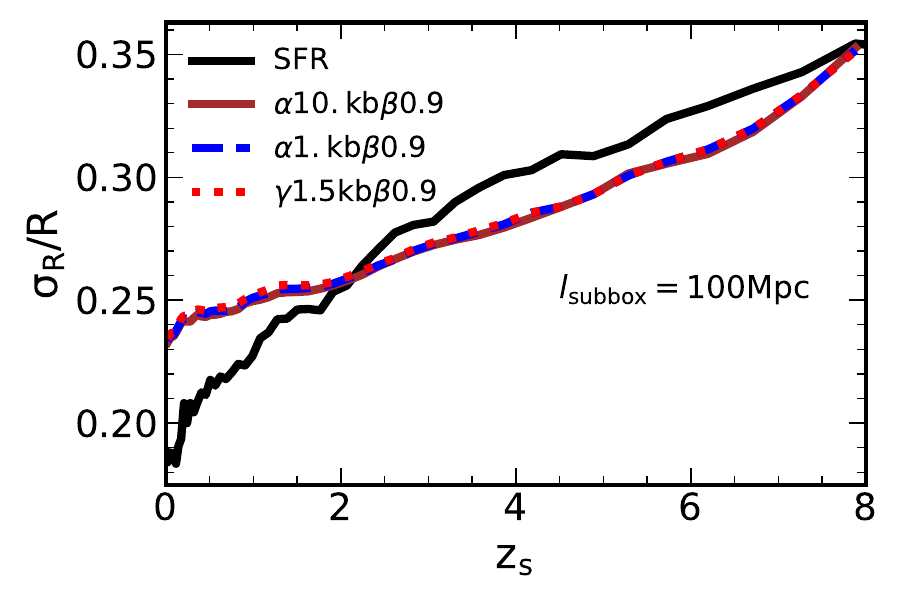}
\includegraphics[width=1.0\columnwidth]{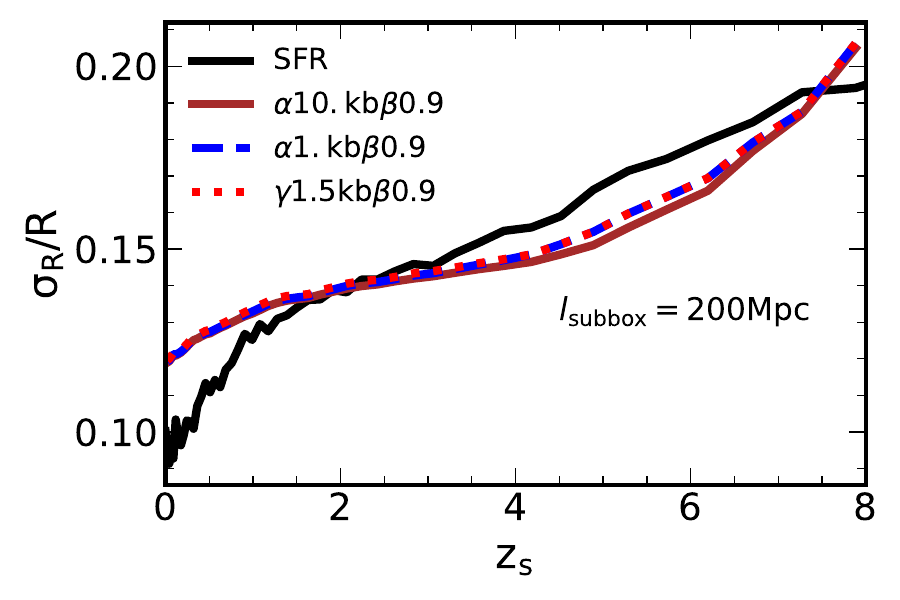}
\caption{ 
The evolution of the relative variance $\sigma(\rm R)/R$ with redshift $z_{\rm s}$ with different sub-box length, i.e., $l_{\rm subbox}=100\rm Mpc$ (upper panel) and $l_{\rm subbox}=200\rm Mpc$ (bottom panel) respectively.  \textbf{The brown solid, blue dashed and red dotted lines} show the results of different parameterized population models, i.e.,  $\boldsymbol{\alpha10.\rm kb\beta0.9}$, $\boldsymbol{\alpha1.\rm kb\beta0.9}$ and $\boldsymbol{\gamma1.5\rm kb\beta0.9}$ respectively. The black solid line shows the relative variance on the SFR.
}
\label{fig:cv}
\end{figure}

To quantify the effect of cosmic variance on  the merger rate density and the detection rate, we define the relative variance of a random variable $X$ with cumulative probability distribution $P(X)$ by the following expression: 
\begin{equation}
    \frac{\sigma_{X}}{X}=\frac{X(P(X)=0.84)-X(P(X)=0.16)}{2X(P(X)=0.5)}. 
\end{equation}
Figure~\ref{fig:cv} further plots the evolution of the relative variance $\sigma_R/R$ with redshift $z_{\rm s}$ for these three models, assuming the sub-box size to be $100$\,Mpc (top panel) and $200$\,Mpc (bottom panel), respectively.
As seen from this figure, the larger the sub-box size, the higher the relative variance at a given redshift. For example, the local relative variance $\sigma_R/\rm R$ is about $\sim 12.0\%$ and $\sim 23.5\%$ for $l_{\rm subbox}\sim100$\,Mpc (top panel) and $200$\,Mpc, respectively. This is easy to understand for the inhomogeneity of star formation decreases with the box size of the simulation, assuming a Gaussian random field. In addition, the evolution and scatter $\sigma_R/\rm R$ resulted from different models are quite similar, though the exact values of $R^{i}_{z_{\rm s}}$ are different. This fact indicates that the estimation on $\sigma_{\rm R}/R$ can be viewed as reasonable and reliable, and their dependence on the choices of different binary population synthesis parameters is weak. 
 
Moreover, the value of $\sigma_{R}/R$ increases with increasing redshift, 
which can be  mainly explained by the star formation history:  the inhomogeneity of star formation is much higher, for voids are more common to exist at early times. As seen from Figure~\ref{fig:cv} that compared with the variance of SFR, $\sigma_{\rm R}/ R$ is suppressed before $z_{\rm s}\sim 2$ for different sub-box size case. This is mainly due to the time-delay from the binary star formation to their merger, which may flatten the difference of different boxes due to the integration effect.

\begin{figure}
\centering
\includegraphics[width=1.0\columnwidth]{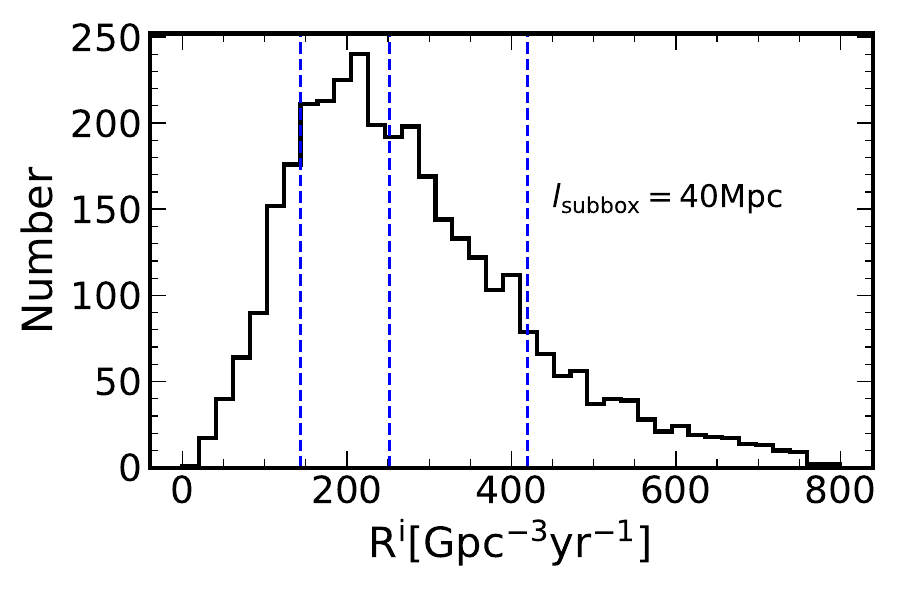}
\caption{ 
The distribution of BNS merger rate density of $15\times15\times15$ realization of sub-boxes, assuming $\boldsymbol{\alpha10.\rm kb\beta0.9}$ BSE model. The length of the sub-box is set to be $40\rm Mpc$, which corresponds to the localization precision of CE networks. The blue dashed lines from left to right show the $16\%$, $50\%$, $84\%$ quantile of the distribution respectively. 
}
\label{fig:dis}
\end{figure}

With next-generation ground-based GW detectors, it is possible to localize BNS mergers within very small area.  For example, with Cosmic Explorer (CE) networks, BNS mergers below $z_{\rm s}\sim 0.2$  are expected to be localized within the sky area of $\Delta\Omega_{\rm s}\sim 0.10\rm deg^2$ and $\Delta d_{\rm L} \sim 20\rm Mpc$ \citep{2018PhRvD..97f4031Z}. In such a small comoving volume, the impact of cosmic variance can be very large. We calculate the distribution of merger rate density with sub-box size of $\rm 40 \rm Mpc$ at $z_{\rm s}=0.2$ and the resulted distribution is shown in Figure~\ref{fig:dis}. It can be seen that the relative variance  $\sigma_{R}/R$ is much larger, i.e., $\sim 55\%$. This result indicates that with future next-generation ground-based GW
detectors, estimating the merger rate density of BNS in
different sky areas may also provide useful information
on the cosmic variance.

With the merger rate density evolution estimated by our ranked-first model $\boldsymbol{\alpha10.\rm kb\beta0.9}$, we may further check the variance on the  detection rate of BNS mergers in O2 run, $\sigma_N/N$, induced by the cosmic variance, following the standard Monte-Carlo procedure. We refer the readers to \citet{2021MNRAS.500.1421Z} and \citet{2023ApJ...953...36C} for details on the generation of mock BNS samples and their SNR calculation. We find that the variance on detection rate is about $\sim 13\%$ for detection threshold $\varrho_{0}=12$ and $\sim 9\%$ for $\varrho_{0}=8$. For GW observation runs post-O2, the variance is much smaller, less than $\lesssim 5\%$. During the writing procedure of this paper, we note that \citet{2024arXiv240600691M} also consider the effects of local cosmic inhomogeneities on the GW event rate, but simply assume the merger rate is proportional to the matter density. Using the peculiar velocities of galaxies from the CosmicFlows (CF) catalogs, they find that the cosmic variance effect on the compact binary coalescence event rate to be at most $\sim 6 \%$, which is almost consistent with our results. Therefore, we conclude that the estimation of local merger rate density from the current observation of LVK is safe by ignoring the cosmic variance, for their very small impact on the detection rate. Here we also noticed that based on the observation of galaxy cluster survey, we find that the density of local volume (within $100\rm Mpc$ radius) is $\sim 30\%$ under-dense from the average across the universe, which indicates that the detection rate of BNS mergers per comoving volume in our local universe may be smaller than the average value across the universe by a factor of $\sim 1-5\%$.

\section{Discussions and conclusions}
\label{sec:con}

In this work, we test the effect of cosmic variance on the merger rate density of binary neutron stars by adopting the galaxy formation model GABE based on the n-body numerical simulation Millennium and the binary neutron star population synthesis model.

We find that for sub-box size of $100$\,Mpc and $200$\,Mpc, the relative variance of merger rate density $\sigma_{R}/R$ at different redshift is about $0.23-0.35$ and $0.13-0.20$, respectively. 
With future powerful next-generation GW detectors, it is anticipated to detect $\sim 10^3-10^4 $  BNS mergers. 
This indicate that one may constrain the cosmic variance via the estimation the BNS merger rate density in different sky areas. Moreover, we estimate the variance on the detection rate of BNS mergers, induced by the cosmic variance. We find that the relative variance of the detection rate is about $\sigma(N)/N\sim 0.05-0.1 $, which is small enough to not significantly impact the inferred merger rate density. As for future next-generation GW detectors, it is possible for one to localize BNS mergers into small sub-boxes. In such small comoving volume, the impact of cosmic variance on the merger rate density and therefore the detection number is much more significant. This offer a possible approach to constrain the cosmic variance by the spatial analysis of BNS mergers.

Note here that as the first estimation, there are many complexities we may need to further taken into account. For example, the absolute value of both $\sigma_{\rm R}/\rm R$ and $\sigma_{N}/N$ may vary slightly due to different settings in the semi-analytical simulations. Moreover, to reduce the cost of computing, we did not run full population synthesis in each sub-box, but use simple description of BNS mergers by the SFR and metallicity to estimate the  $\sigma_{\rm R}/\rm R$. This may also lead to slight difference on the results, though do not affect our main claim.  

One may also conduct the parallel analysis to the stellar binary black hole mergers to check the effect of cosmic variance on their merger rate density. Besides, it is also possible to find a small sub-box very similar with our local observation in this model, which may provide interesting insights to the estimation of local merger rate density of compact binaries. We defer these aspects to future work. 

\section*{acknowledgement}
Zhiwei Chen thanks the daily-AstroCoffee organized in National Astronomical Observatories of China, and the speakers and audiences therein for their helpful discussions.  This work is partly supported by the Strategic Priority Research Program of the Chinese Academy of Sciences (Grant no. XDB0550300), the National Natural Science Foundation of China (Grant nos. 12273050, 11991052), and the National Key Program for Science and Technology Research and Development (Grant nos. 2020YFC2201400 and 2022YFC2205201).

\bibliographystyle{aasjournal}
\bibliography{ref.bib}

\end{document}